# Photocatalytic Carbon Dioxide Methanation by High-Entropy Oxides: Significance of Work Function

Jacqueline Hidalgo-Jiménez[1,2], Taner Akbay[3], Xavier Sauvage[4], Tatsumi Ishihara[1,5] and Kaveh Edalati[1,5,*]

[1] WPI, International Institute for Carbon Neutral Energy Research (WPI-I2CNER), Kyushu University, Fukuoka 819-0395, Japan
[2] Department of Automotive Science, Kyushu University, Fukuoka, Japan
[3] Materials Science and Nanotechnology Engineering, Yeditepe University, Istanbul, Turkey
[4] Univ Rouen Normandie, INSA Rouen Normandie, CNRS, Groupe de Physique des Matériaux, UMR6634, 76000 Rouen, France
[5] Mitsui Chemicals, Inc. - Carbon Neutral Research Center (MCI-CNRC), Kyushu University, Fukuoka, Japan

Methane ($CH_4$) formation from photocatalytic carbon dioxide ($CO_2$) conversion in water is currently of interest because methane is a fuel, and it can also be transformed into other useful hydrocarbons. However, achieving high selectivity to produce methane remains a challenge because of the large number of contributing electrons (eight) in methanation. High-entropy oxides present a new pathway to tune the catalyst selectivity by arranging various cations in the lattice. This study aims to clarify the selectivity for methane formation in high-entropy photocatalysts containing hybrid $d^0 + d^{10}$ orbital configuration. Several oxides are designed and synthesized which have a base of 3-4 cations with $d^0$ orbital configuration (titanium and zirconium with a valence of 4, and niobium and tantalum with a valence of 5) and incorporate 1-2 elements with $d^{10}$ orbital configuration (zinc, gallium, indium, bismuth and copper). Results demonstrate that adding elements with a $d^{10}$ electronic configuration is effective for methane formation, while the selectivity toward methanation is enhanced by increasing the work function of the $d^{10}$ cations. Selectivity levels over 50% are achieved using these oxides, suggesting a potential strategy for designing new catalysts for methanation.

*Corresponding author: (E-mail: kaveh.edalati@kyudai.jp; Tel/Fax: +81 92 802 6744)

## 1. Introduction

One of the major contaminants produced by industry is $CO_2$ due to the use of fossil fuels. The swift rise in the atmospheric $CO_2$ concentration is directly related to environmental issues such as climate change [1,2]. Reducing the atmospheric $CO_2$ concentration is necessary to improve the continuing rise in global temperature, and thus, exploring new clean technologies plays a crucial role in dealing with $CO_2$ emissions and global warming. $CO_2$ removal is currently possible thanks to carbon capture and storage technologies [3]; however, there is a need to transform the captured $CO_2$ into useful products to develop a carbon-neutral society [2,4,5].

Solar-driven technologies, such as photocatalysis, to transform $CO_2$ into useful compounds by catalytic reactions are considered a promising solution [2,4,6]. Photocatalysis was first introduced in the 1970s for the water-splitting reaction [7], while the process has been used for performing different reactions for the past five decades [2,8]. The process consists of the excitation of a semiconductor by using light, followed by charge separation, and finally, reactions between charge carriers and the molecules adsorbed on the surface [9]. During $CO_2$ photoreduction, multiple products can be obtained, such as CO, $CH_4$, $CH_3OH$ and HCOOH [6,10]. Particularly, the $CH_4$ production (or methanation), is one of the most promising routes of the hydrogenation of $CO_2$. The produced $CH_4$ can be utilized for power generation as a fuel, or for the production of other hydrocarbons such as methanol or biodiesel [11]. In this process, the first step is hydrogen production which is followed by the Sabatier reaction at 298 K [5,10,11].

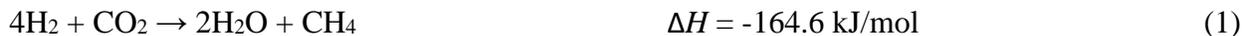

$$4H_2 + CO_2 \rightarrow 2H_2O + CH_4 \qquad \Delta H = -164.6 \text{ kJ/mol} \qquad (1)$$

This is a thermodynamically favorable and exothermic reaction, but it requires eight electrons, which leads to kinetic limitations [12]. Generally, increasing the temperature can improve $CO_2$ conversion as it contributes to overcoming the energy barriers for methanation. In a photocatalytic reactor, the temperature is often room temperature and only light is utilized as an excitation source of electrons in the catalyst [5]. Therefore, the reaction efficiency is dependent on the photoactivity and selectivity of photocatalysts. Many semiconductors have been tested for this application, such as $TiO_2$ [13,14], $BiVO_4$ [15], CuO [16], and various other oxides and oxynitrides. Due to the difficulties of the methanation, many of these catalysts need to be improved for the reaction by proper strategies. Different strategies, like defect engineering [2,8,15], dopant addition [17] and cocatalyst modification [18], have been used in this regard. Recently, high-entropy catalysts were employed for $CO_2$ reduction which can accommodate several strategies in a single material [19–26].

High-entropy materials, formed by five or more principal elements, provide several advantages for catalytic applications [25]. First, the variety of elements utilized allows the manipulation of the chemical composition flexibility to obtain better properties. For example, recent studies succeeded in mixing cations with $d^0 + d^{10}$ electronic configuration which is beneficial for catalytic water splitting [27,28]. The mixed electronic configuration in hybrid structures was reported to be effective for water splitting. Second, the random distribution of the elements allows multiple active sites with different performances for different reactions. Third, the high-entropy nature results in reducing Gibbs free energy and enhanced stability [10,29]. Fourth, the combination of five cations in HEOs generates a severe distortion in the crystal structure [30], expecting to generate internal electric fields, promote the separation of photo-induced carriers and affect $CO_2$ adsorption configuration and selectivity [31,32]. In addition, the appropriate compositional design and synthesis are expected to lead to better chemical selectivity for hard-to-perform catalytic reactions such as methanation, although there have been no attempts in this regard so far.

In this study, six new high-entropy oxides (HEOs) were synthesized using the mixed $d^0 + d^{10}$ electronic configuration concept ($d^0$ elements: titanium, zirconium, niobium and tantalum; $d^{10}$ elements: zinc, gallium, copper and bismuth) and their selectivity for methanation was examined. The addition of $d^{10}$ elements enhances the selectivity for methanation, while the selectivity is higher for cations with a higher work function. This study not only introduces six novel HEOs with mixed $d^0 + d^{10}$ electronic configuration but, more importantly, highlights a potential approach to enhance the selectivity toward methane formation using work function.

## 2. Methodology
### 2.1. Synthesis

A severe plastic deformation technique of high-pressure torsion (HPT), combined with annealing, was implemented to synthesize HEOs. Initially, a combination of binary oxides of the $d^0$ elements (titanium oxide Sigma Aldrich 99.8%, zirconium oxide Kojundo 97%, niobium oxide Kojundo 99% and tantalum oxide Kojundo 99.9%) were mixed with different oxides of $d^{10}$ elements (zinc oxide Kanto Chemical Co. Inc. 99%, gallium oxide Kojundo 99.99%, copper oxide Sigma Aldrich MQ100, indium oxide Kojundo 99.9% and bismuth oxide Kojundo 99.9%) to generate mixtures with overall compositions of $TiZrNbTaGaO_{10.5}$, $TiZrNbTaInO_{10.5}$, $TiZrNbTaBiO_{10.5}$, $TiZrNbTaZnO_{10}$, $TiNbTaGaBiO_{10}$ and $TiNbTaCuZnO_9$. In addition, one sample with a composition of $TiZrHfNbTaO_{11}$, containing only elements with the $d^0$ electronic configuration, was synthesized for comparison. The mixture of the binary oxides was first mixed by hand in acetone for 30 min employing a mortar, then compacted in the form of pellets with 0.8 mm thickness and 5 mm radius to be treated by HPT. The HPT process was performed under 6 GPa pressure, for 3 rotations at room temperature with a rate of one rotation per minute. Since mechanical strain in HPT depends on the distance from the disc center (low strain in the center and high strain in the edge), the HPT-treated samples were crushed, and mixed and treated by the HPT process for an extra 3 turns to have a better homogeneity. The resulting discs were heated to 1373 K and kept at this temperature for one day (24 h). Then, an additional HPT process for another 3 turns was performed under the same conditions, and the sample was calcinated again at 1373 K for 24 h to have a complete homogenization of the elements at the atomic scale.

### 2.2. Characterization

The HEOs were examined using different techniques as follows.

First, to determine the crystallographic features of the different materials, X-ray diffraction (XRD) was performed utilizing a Cu Kα X-ray source. XRD profiles were evaluated utilizing the Rietveld refinement and Halder-Wegner methods to recognize the lattice parameters and measure crystallite sizes. Furthermore, the characteristic vibrational modes of all compositions were examined by Raman spectroscopy using a laser lamp with 532 nm wavelength.

Second, the oxidation status and surface composition of oxides were examined utilizing X-ray photoelectron spectroscopy (XPS) with an Al Kα source.

Third, the composition and microstructural features of the materials were tested utilizing scanning electron microscopy (SEM) under 15 keV voltage by employing energy-dispersive X-ray spectroscopy (EDS). Microstructure was also tested using transmission electron microscopy (TEM) under 200 kV voltage by employing high-resolution images and selected area electron diffraction (SAED). The chemical composition homogeneity at the nanometer scale was further analyzed by EDS in scanning-transmission electron microscopy (STEM). The samples for TEM/STEM were prepared by crushing powders and dispersing them onto carbon-covered copper grids with 1.5 mm radius.

Fourth, bandgaps were obtained by light absorbance measurements utilizing UV-vis spectroscopy followed by the Kubelka-Munk analysis. This bandgap calculation was combined with XPS results in a low binding energy range (for determining the valence band maximum energy) to estimate the electronic structure of the new HEOs.

Fifth, the electron-hole radiative recombination was tested utilizing photoluminescence spectroscopy and a 325 nm laser source. Moreover, the oxygen vacancy presence, which can function as active sites or charge carrier recombination sites, was examined by electron spin resonance (ESR) using a microwave with 9468.8 MHz frequency.

## 2.3. Photocatalysis

The catalytic performance of the HEOs was examined for methane formation using $CO_2$ conversion reactions in water. The reaction was performed in a quartz photoreactor with a volume of 858 mL and a continuous flow of $CO_2$. On top of the reactor, a $CO_2$ gas input was connected with a 30 mL/min flow rate to bubble the gas into the aqueous solution of the reactor. In addition, a second output pipe connected the reactor with two gas chromatographs (GC). The first GC (GF-8A of Shimadzu) was used for oxygen and hydrogen detection, and the second GC (GC-4000 coupled with MT 221 methanizer of GL Science) was utilized to analyze the $CO_2$, CO and $CH_4$ concentrations. The reactor contained a solution of catalyst (100 mg), $NaHCO_3$ (4.2 g) and water (500 mL). To ensure the suspension of the particles, this solution was continuously stirred utilizing a magnetic stirrer with a stirring rate of 420 rotations per minute. The temperature of the reactor was regulated at 293 K utilizing a water chiller. The irradiation was conducted utilizing the full arc of a 400 W high-pressure mercury lamp without using any filters. The mercury lamp was inserted in a hole in the inner space of the photoreactor, resulting in 1.4 W/cm$^2$ light intensity on the reactants. Before the irradiation of the sample, $CO_2$ gas was injected for 1 h to verify the absence of air or reaction products in the photoreactor. Moreover, a blank test was carried out under illumination for 24 h and without catalyst addition to confirm the absence of reactions in the reactor.

## 3. Results
### 3.1. Crystal structures

After the completion of the synthesis process, the crystal structure of all the HEOs was tested utilizing XRD as shown in Fig. 1. Fig 1(a) shows the final crystal structure and the corresponding Rietveld refinement of $TiZrNbTaGaO_{10.5}$. This sample corresponds to a dual-phase material with an orthorhombic (Pbcn space group) crystal structure and a monoclinic (C2/m space group) crystal structure. Rietveld analysis revealed that the orthorhombic crystal structure occupies 88 wt% of the synthesized catalyst while the fraction of monoclinic is only 12 wt%. Fig 1(b) shows the examination of $TiZrNbTaInO_{10.5}$, in which two phases are observed: (1) 83 wt% of monoclinic with a P2/c space group and (2) 17 wt% of tetragonal with P4bm space group. For the third HEO, $TiZrNbTaBiO_{10.5}$, the XRD profile is illustrated in Fig 1 (c), suggesting the presence of two phases: 95 wt% of cubic (Fd-3m space group) and 5 wt% of tetragonal (P4$_2$/nmc space group). The fourth sample $TiZrNbTaZnO_{10}$ has a single monoclinic phase with a space group of P2/c (Fig. 1(d)). Fig. 1(e) shows the analysis of the crystal structure for $TiNbTaGaBiO_{10}$ which has a dual-phase structure. A tetragonal phase (P4$_2$/mnm space group) and a cubic phase (Fd-3m space group) were observed with fractions of 48 wt% and 52 wt%, respectively. Last of all, Fig. 1(f) shows the Rietveld analysis for $TiNbTaCuZnO_9$ which has a single tetragonal crystal structure and P4$_2$/mnm space group. Table 1 shows the crystallographic information for all these HEOs with $d^0 + d^{10}$ electronic configuration as well as TiZrHfNbTaO$_{11}$ with only $d^0$ electronic configuration with XRD patterns reported in an earlier publication [33]. Taken altogether, XRD profiles confirm the successful production of several single- and two-phase HEOs.

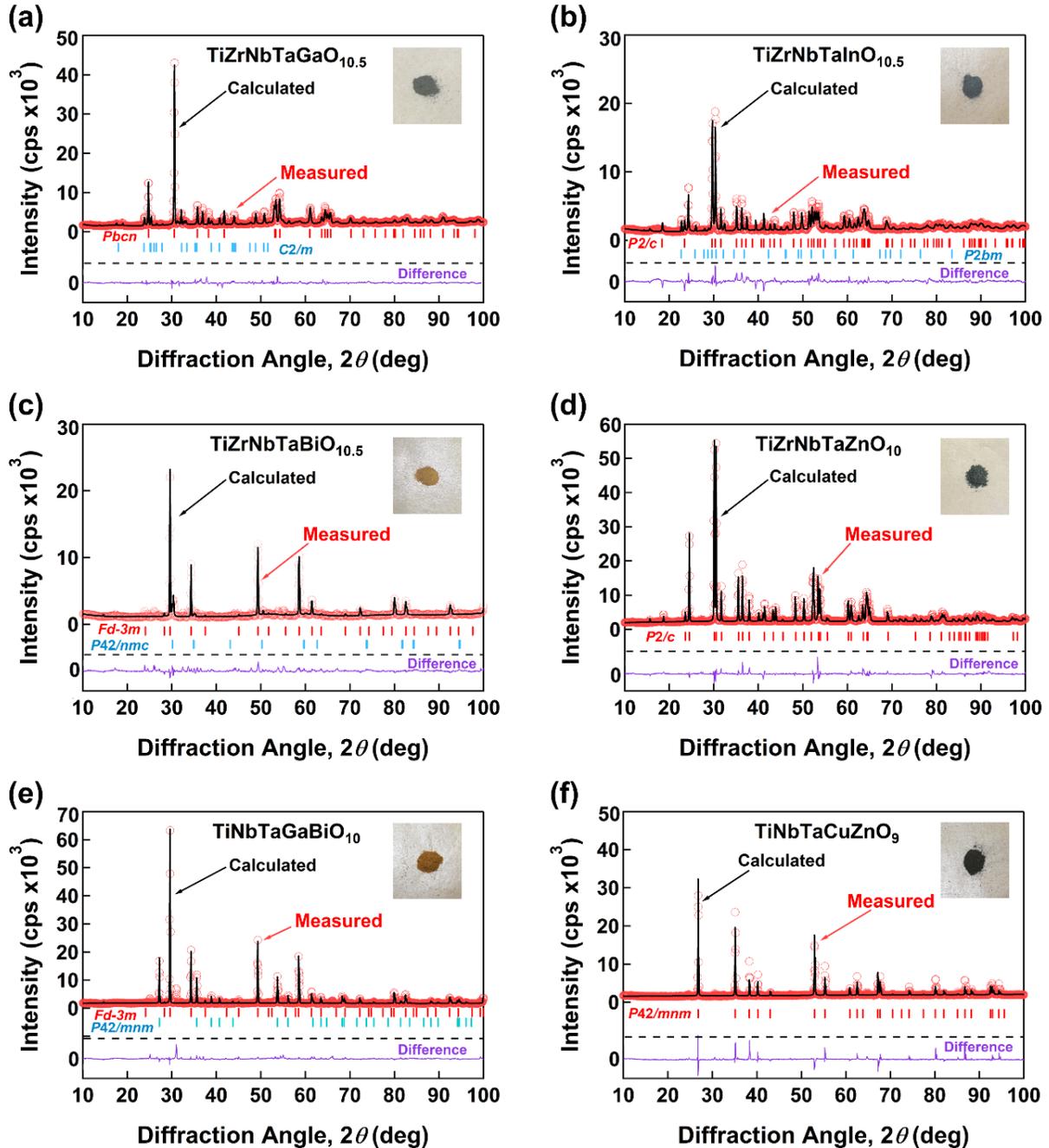

Fig. 1. Successful synthesis of high-entropy oxides with single and dual phases and $d^0 + d^{10}$ electronic configuration. XRD patterns and corresponding Rietveld analysis of (a) TiZrNbTaGaO$_{10.5}$, (b) TiZrNbTaInO$_{10.5}$, (c) TiZrNbTaBiO$_{10.5}$, (d) TiZrNbTaZnO$_{10}$, (e) TiNbTaGaBiO$_{10}$ and (f) TiNbTaCuZnO$_9$. Inset: appearance of samples.

Table 1. Lattice parameters achieved by Rietveld method for high-entropy oxides TiZrNbTaGaO$_{10.5}$, TiZrNbTaInO$_{10.5}$, TiZrNbTaBiO$_{10.5}$, TiZrNbTaZnO$_{10}$, TiNbTaGaBiO$_{10}$ and TiNbTaCuZnO$_9$.

| Composition | Space Group | | $a$ (Å) | $b$ (Å) | $c$ (Å) | $\alpha$ (°) | $\beta$ (°) | $\gamma$ (°) | Fraction (wt%) |
|---|---|---|---|---|---|---|---|---|---|
| **TiZrNbTaGaO$_{10.5}$** | Orthorhombic | Pbcn | 4.700 (4) | 5.569 (9) | 5.018 (1) | 90 | 90 | 90 | 88 |
| | Monoclinic | C2/m | 15.843 (9) | 3.853 (2) | 17.946 (1) | 90 | 90 | 90 | 12 |
| **TiZrNbTaInO$_{10.5}$** | Monoclinic | P2/c | 4.809 (7) | 5.658 (8) | 5.112 (7) | 90 | 91.8 | 90 | 83 |
| | Tetragonal | P2bm | 12.388 (17) | 12.388 (17) | 3.891 (13) | 90 | 90 | 90 | 17 |
| **TiZrNbTaBiO$_{10.5}$** | Cubic | Fm-3m | 10.4545 (1) | 10.4545 (1) | 10.4545(1) | 90 | 90 | 90 | 95 |
| | Tetragonal | P4$_2$/nmc | 3.596 (7) | 3.596 (7) | 5.115 (7) | 90 | 90 | 90 | 5 |
| **TiZrNbTaZnO$_{10}$** | Monoclinic | P2/c | 4.740 (1) | 5.646 (1) | 5.055 (1) | 90 | 91.1 | 90 | 100 |
| **TiNbTaGaBiO$_{10}$** | Tetragonal | P4$_2$/mnm | 4.630 (1) | 4.630 (1) | 3.004 (1) | 90 | 90 | 90 | 52 |
| | Cubic | Fm-3m | 10.453 (1) | 10.453 (1) | 10.453 (1) | 90 | 90 | 90 | 48 |
| **TiNbTaCuZnO$_9$.** | Tetragonal | P4$_2$/mnm | 4.696 (3) | 4.696 (3) | 3.043 (2) | 90 | 90 | 90 | 100 |
| **TiZrHfNbTaO$_{11}$** | Monoclinic | A2/m | 11.870 (7) | 3.820 (15) | 20.550 (11) | 90 | 119.9 | 90 | 60 |
| | Orthorhombic | Ima2 | 45.970 (18) | 4.620 (23) | 5.120 (26) | 90 | 90 | 90 | 40 |

### 3.2. Microstructures

The chemical composition homogeneity of the samples was evaluated using two methods: SEM-EDS at a microscopic scale and STEM-EDS at a nanoscopic scale. Fig 2 shows the distribution of different elements in the HEOs at the microscopic scale. Dual-phase samples are confirmed to have regions rich in one or two elements. For example, TiZrNbTaGaO$_{10.5}$ shows a main phase containing all the elements and a secondary Ta-Nb-rich phase. Similarly, TiZrNbTaInO$_{10.5}$ shows a secondary phase rich in tantalum and niobium and a main phase containing all the elements. TiZrNbTaBiO$_{10.5}$ has a Bi-rich phase and another one with a more homogeneous distribution of all the elements. The last dual-phase sample is TiNbTaGaBiO$_{10}$ in which the separation of the two phases is the most evident, since it shows a tendency for gallium and bismuth to be partitioned. The distribution of titanium, tantalum and niobium in this HEO is more homogeneous, although there is a certain preference for titanium to be in the Ga-rich phase. On the other hand, samples with single crystal structures, TiZrNbTaZnO$_{10}$ and TiNbTaCuZnO$_9$, show homogeneous distribution of the elements in all the samples. The compositions measured by SEM-EDS analysis are summarized in Table 2, showing a reasonable consistency with the nominal equiatomic compositions.

To determine the composition of each phase in dual-phase HEOs, the comparison of the STEM-EDS results and the high-resolution TEM images and SAED patterns obtained in the same regions was conducted. Fig. 3 shows the elemental distribution for the HEOs at the nanometric scale achieved by STEM-EDS, and Fig. 4 shows some representative high-resolution TEM and SAED patterns. Detailed analysis of these two figures gives information about the composition of each phase in each HEO. For example, in TiZrNbTaGaO$_{10.5}$, the SAED pattern (Fig. 4a) suggests the space group C2/m and the [102] orientation in the Ti-Nb-rich region. A similar analysis performed in the regions containing all the elements confirmed the orthorhombic crystalline phase with the space group of Pbcn for the matrix. Fig. 4b show the high-resolution TEM analysis and corresponding SAED pattern for the monoclinic phase of TiZrNbTaInO$_{10.5}$. The analysis was performed in a region containing all the elements, confirming the monoclinic crystal structure as the main phase of TiZrNbTaInO$_{10.5}$. For TiZrNbTaBiO$_{10.5}$, the selected region shown in Fig. 4c was obtained from a Bi-rich phase, corresponding to the cubic phase with Fm-3m space group in [100] orientation. Fig. 4d, corresponding to the TEM analysis of TiZrNbTaZnO$_{10}$, confirms the monoclinic crystal structure of the sample. In addition, STEM-EDS results confirm the uniform distribution of elements at the nanometric level in this single-phase oxide. For TiNbTaGaBiO$_{10}$, both phases are in a similar fraction and the Ga-Ti-rich phase can be assigned to the tetragonal crystal structure and the Bi-rich phase to the cubic crystal structure (Fig. 4e was obtained from the Ga-Ti rich phase). Finally, for TiNbTaCuZnO$_9$, the tetragonal single-crystal structure was verified as illustrated in Fig. 4f. Taken altogether, a combination of high-resolution TEM images and STEM-EDS mappings confirms the good atomic-scale mixing of in each phase of the HEOs.

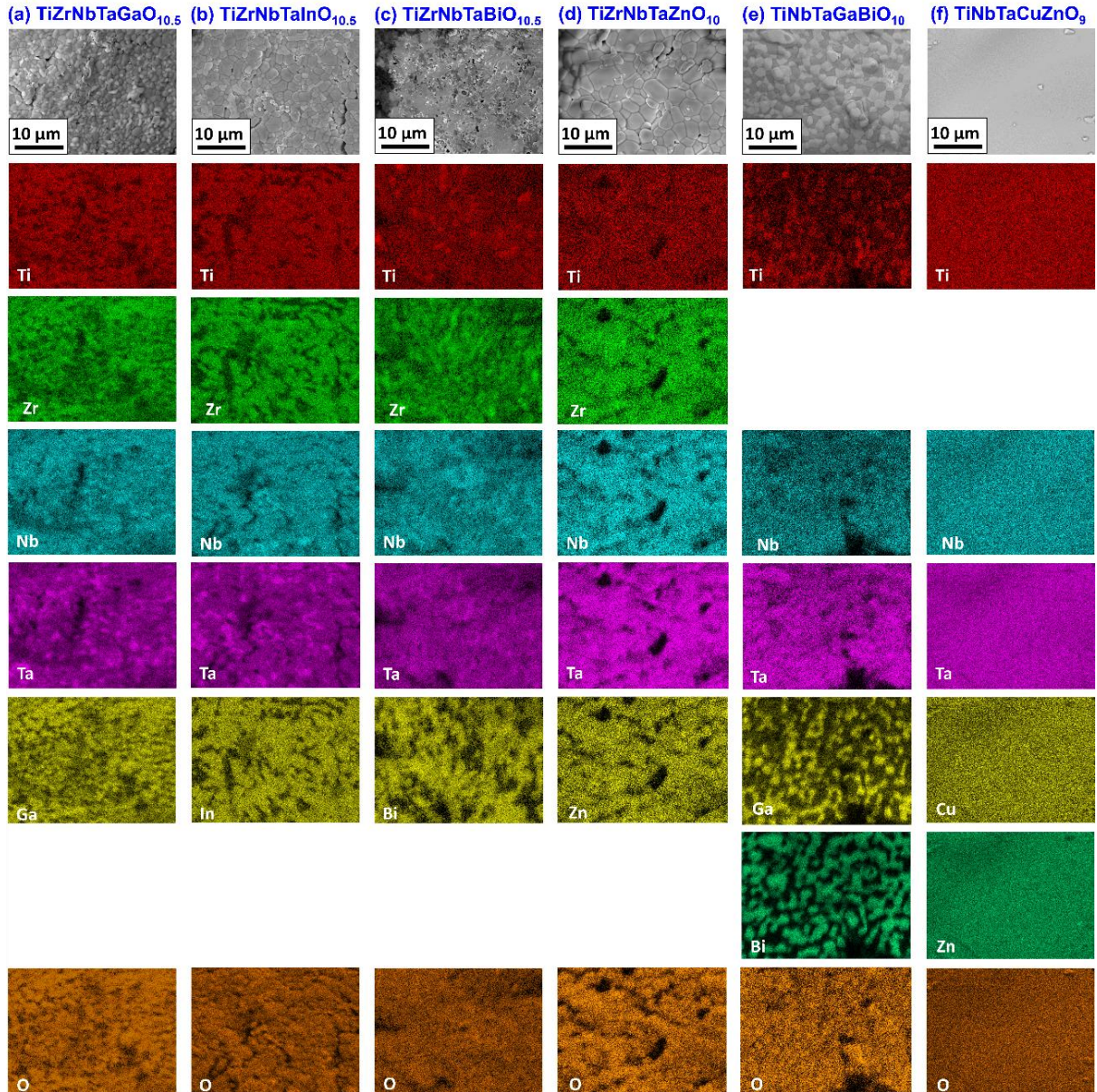

Fig. 2. Distribution of titanium, zirconium, niobium, tantalum, gallium, indium, bismuth, zinc, copper and oxygen at micrometer scale in high-entropy oxides. SEM-EDS mapping for (a) TiZrNbTaGaO$_{10.5}$, (b) TiZrNbTaInO$_{10.5}$, (c) TiZrNbTaBiO$_{10.5}$, (d) TiZrNbTaZnO$_{10}$, (e) TiNbTaGaBiO$_{10}$ and (f) TiNbTaCuZnO$_9$.

Table 2. Compositions of high-entropy oxides $TiZrNbTaGaO_{10.5}$, $TiZrNbTaInO_{10.5}$, $TiZrNbTaBiO_{10.5}$, $TiZrNbTaZnO_{10}$, $TiNbTaGaBiO_{10}$ and $TiNbTaCuZnO_9$ measured by SEM-EDS.

| Element | $TiZrNbTa$ $GaO_{10.5}$ | $TiZrNbTa$ $InO_{10.5}$ | $TiZrNbTaBi$ $O_{10.5}$ | $TiZrNbTa$ $ZnO_{10}$ | $TiNbTaGaBi$ $O_{10.5}$ | $TiNbTaCu$ $ZnO_9$ | $TiZrHfNb$ $TaO_{11}$ |
|---|---|---|---|---|---|---|---|
| Ti | 3.2 ± 1.5 | 4.5±1.1 | 5.8±1.0 | 4.6±0.4 | 5.2±1.9 | 8.6±1.9 | 6.8±1.5 |
| Zr | 4.0±0.7 | 5.7±0.9 | 3.1±1.3 | 5.2±0.4 | | | 4.6±1.4 |
| Hf | | | | | | | 6.7±1.0 |
| Nb | 4.7±0.8 | 5.2±1.8 | 8.0±1.5 | 5.2±0.5 | 5.8±0.5 | 6.6±0.7 | 5.0±1.3 |
| Ta | 4.6±0.7 | 5.4±1.7 | 7.0±0.5 | 5.5±0.5 | 6.4±0.4 | 7.8±1.2 | 4.9±0.5 |
| Ga | 4.9±0.1 | | | | 5.8±3.0 | | |
| In | | 5.3±1.7 | | | | | |
| Bi | | | 3.1±3.8 | | 5.8±3.8 | | |
| Zn | | | | 6.8±0.1 | | 6.8±1.0 | |
| Cu | | | | | | 4.3±2.4 | |
| O | 78.7±3.8 | 73.9±5.4 | 73.0±2.4 | 72.3±1.7 | 71.0±1.5 | 66.0±1.5 | 72.0±2.5 |

### 3.3. Oxygen Vacancies

Oxygen vacancies are common features in high-entropy materials and their presence might affect their optical properties. The presence of this defect was analyzed by ESR examinations as shown in Fig. 5. ESR results indicate the existence of a dual peak with turning points close to $g = 2.002$ and $2.01$, suggesting the presence of vacancy-type defects in their structure. The value of $g$ in the range of 2.002 to 2.004 corresponds to $F^0$ or $V_O$ oxygen vacancies with two trapped electrons, while $g = 2.01$ usually corresponds to paramagnetic oxygen vacancies in strained structure [2,34]. The peaks observed at $g$ values smaller than 2 should be due to cation centers such as paramagnetic $Ti^{3+}$ [34–36]. The oxygen vacancy formation energy in HEOs is strongly dependent on the local chemical environment due to the structural strain and distortions in the structures [29,37]. Therefore, the concentration of these defects is also dependent on their chemical composition. ESR results suggest that $TiZrNbTaGaO_{10.5}$, $TiZrNbTaZnO_{10}$ and $TiNbTaCuZnO_9$ have their peak intensities higher compared to that of other oxides, although a higher peak intensity does not necessarily mean a higher oxygen vacancy concentration. It should be noted that shoulders to high binding energies appear in XPS plots for oxygen, as shown in Fig 5 (b) and the supporting information. These shoulders provide evidence for the formation of oxygen vacancies and highly defective surfaces in these HEOs. However, the quantification of oxygen vacancies by peak deconvolution cannot reliably be done due to the presence of various cations and the overlap of oxygen vacancy peaks and absorbed -OH peaks. Oxygen vacancies have an essential function in photocatalysis as they can act as active sites or recombination centers dependent on their location (surface or bulk) and their concentration in the material [2,38,39].

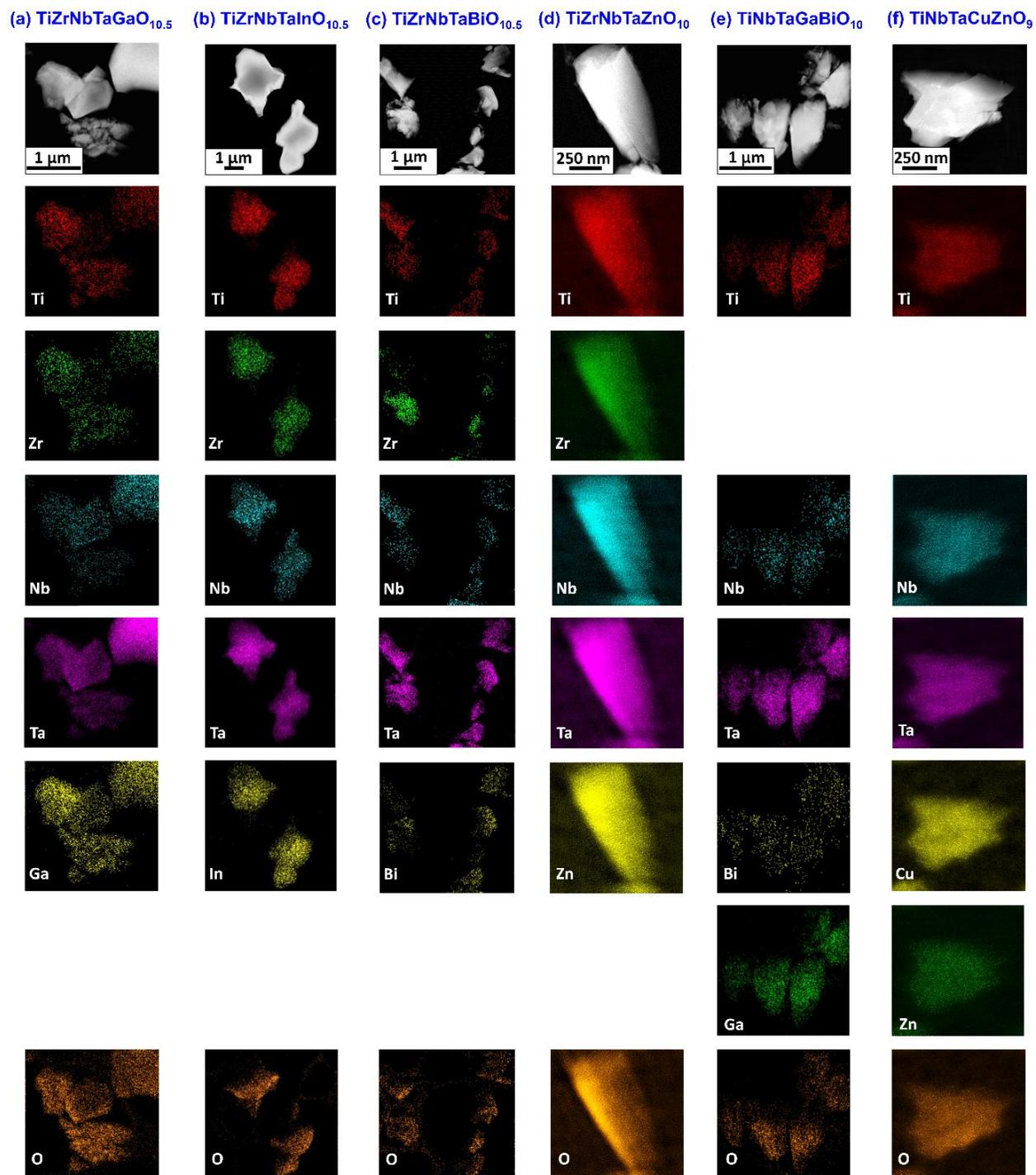

Fig. 3. Distribution of various cations and oxygen at nanometer scale in high-entropy oxides. SEM-EDS mapping for (a) TiZrNbTaGaO$_{10.5}$, (b) TiZrNbTaInO$_{10.5}$, (c) TiZrNbTaBiO$_{10.5}$, (d) TiZrNbTaZnO$_{10}$, (e) TiNbTaGaBiO$_{10}$ and (f) TiNbTaCuZnO$_9$.

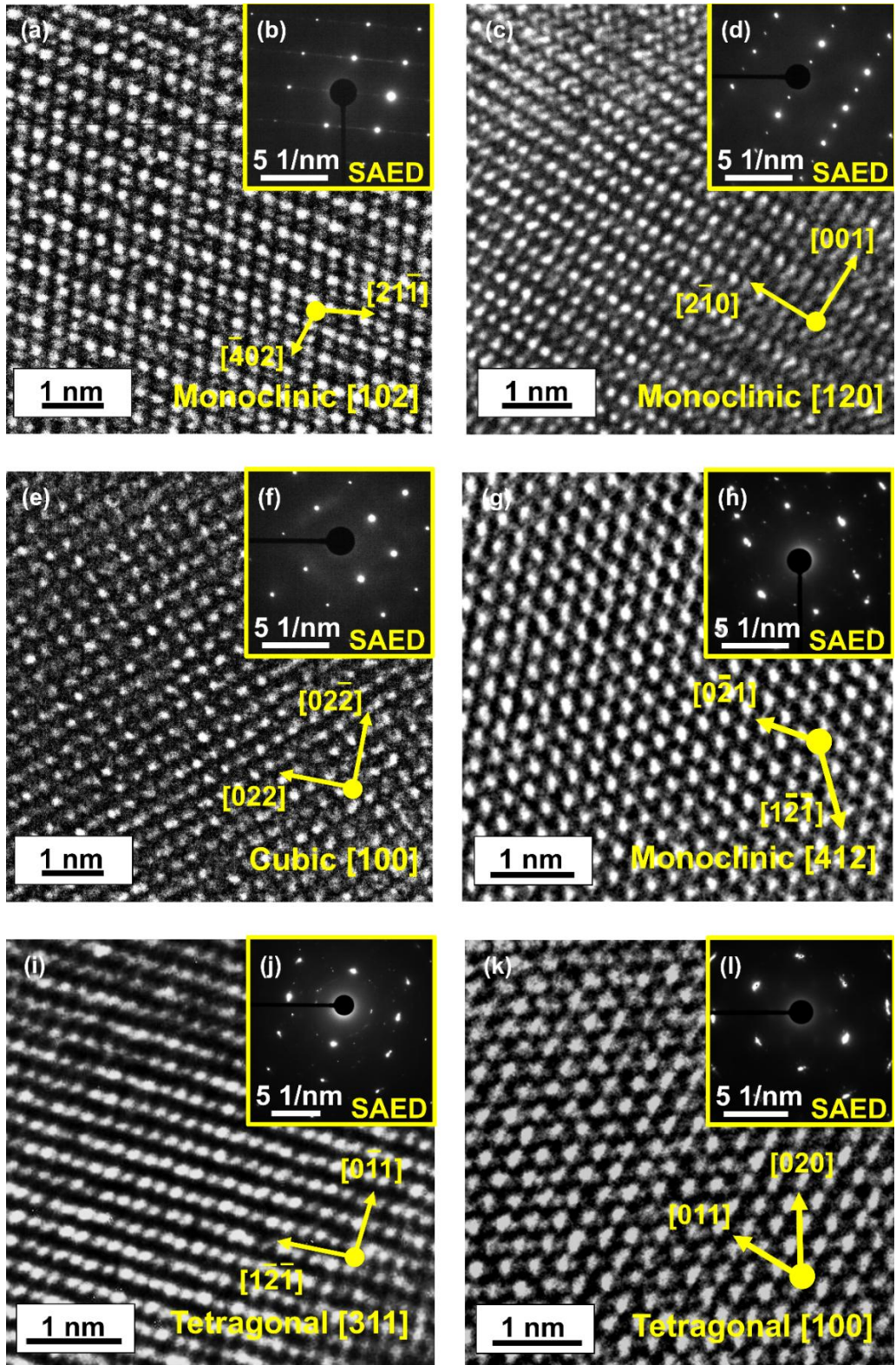

Fig. 4. Atomic-scale images of high-entropy oxides. High-resolution TEM micrographs and SAED patterns obtained in the same area for (a) TiZrNbTaGaO$_{10.5}$, (b) TiZrNbTaInO$_{10.5}$, (c) TiZrNbTaBiO$_{10.5}$, (d) TiZrNbTaZnO$_{10}$, (e) TiNbTaGaBiO$_{10}$ and (f) TiNbTaCuZnO$_9$.

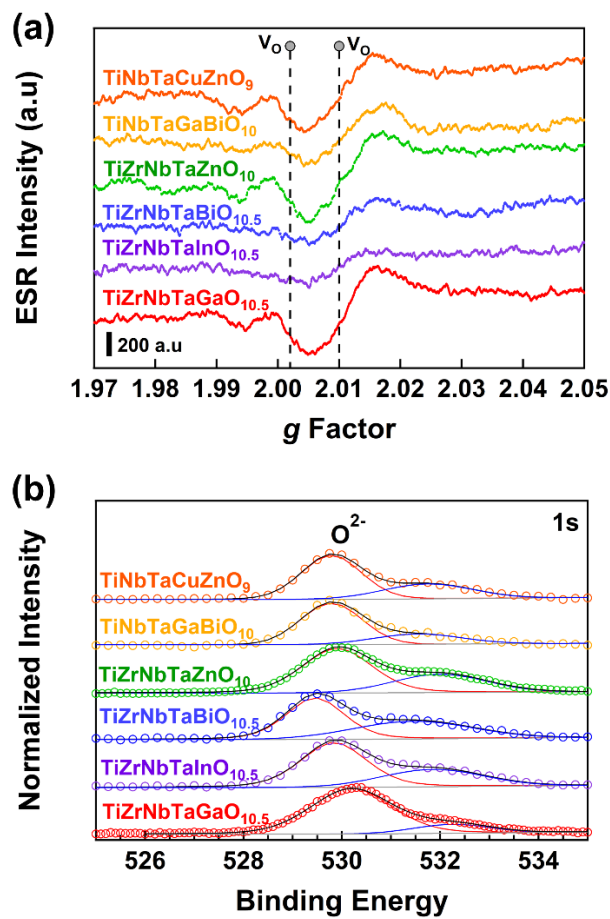

Fig. 5. Formation of oxygen vacancy defects in high-entropy oxides. (a) ESR and (b) XPS O 1s spectra and corresponding peak deconvolution for TiZrNbTaGaO$_{10.5}$, TiZrNbTaInO$_{10.5}$, TiZrNbTaBiO$_{10.5}$, TiZrNbTaZnO$_{10}$, TiNbTaGaBiO$_{10}$ and TiNbTaCuZnO$_9$.

### 3.4. Optical Properties

Optical properties were analyzed to have a better understanding of these HEOs as photocatalysts. The examination of these properties was performed by Raman spectra, UV-vis spectra, and XPS as illustrated in Fig 6. Fig 6a shows the results of Raman spectroscopy for all the synthesized HEOs. Note that such Raman spectra for each sample were achieved in three different positions which showed high similarity, confirming the homogeneity of the sample within the detection limits of Raman spectroscopy. Table 3 shows the Raman shifts for each sample. Based on the reported information in different publications, it is possible to identify the vibrational modes related to different peaks. For example, the peaks located in the wavenumber less than 250 cm$^{-1}$ are usually related to lattice vibration resulting from the oxygen-cation vibrational modes [40], while such vibrations are consistently visible for all the samples. The vibrational modes observed in 250-400 cm$^{-1}$ range are usually related either to oxygen-cation-oxygen bending vibration [40] or to Nb/Ta-O [41], Zr-O [42] and Zn-O [41] asymmetric stretching. For TiZrNbTaInO$_{10.5}$, the existence of the predominant peak at 366 cm$^{-1}$ is characteristic of the stretching vibration of InO$_6$ octahedrons and it was reported for indium oxide [43]. The peaks observed in the 630-670 cm$^{-1}$ range should correspond to TiO$_6$ octahedra and Ti-O stretching vibrational modes [42,44]. This latter feature appears in all samples as titanium is present in all compositions. Finally, all the samples have peaks in the range after 800 cm$^{-1}$, which can be attributed to TaO$_6$ or NbO$_6$ octahedra [41,42,45,46].

Table 3. Modes for Raman Shift observed in high-entropy oxides TiZrNbTaGaO$_{10.5}$, TiZrNbTaInO$_{10.5}$, TiZrNbTaBiO$_{10.5}$, TiZrNbTaZnO$_{10}$, TiNbTaGaBiO$_{10}$ and TiNbTaCuZnO$_9$.

| Raman Shift | | | | | | Vibrational mode | Ref |
|---|---|---|---|---|---|---|---|
| TiZrNbTa GaO$_{10.5}$ | TiZrNbTaIn O$_{10.5}$ | TiZrNbTaBi O$_{10.5}$ | TiZrNbTa ZnO$_{10}$ | TiNbTaGaBi O$_{10}$ | TiNbTaCu ZnO$_9$ | | |
| 123 | 128 | 133 | 130 | 150 | 131 | Lattice | [40] |
|  |  |  | 185 |  |  | Zr-O stretch | [42] |
|  |  |  | 249 |  | 259 | O-cation-O bend | [40] |
| 264 | 274 | 276 | 280 | 273 |  | Nb/Ta-O | [41] |
|  |  |  | 307 |  |  | Zn-O symmetric stretch | [41] |
|  |  | 366 | 355 |  |  | O-Zr-O bend | [42] |
|  | 366 |  |  |  |  | In-O | [43] |
|  |  |  | 405 |  |  | Zr-O stretch | [42] |
| 421 | 421 |  |  |  | 388 | Ti-O | [47] |
|  |  | 435 |  | 447 | 431 | | |
|  |  |  | 465 |  |  | O-Zr-O bend | [42] |
|  |  |  | 511 |  |  | Zr-O | [42] |
| 602 | 574 | 617 | 593 | 613 |  | Nb/Ta-O asymmetric vibration | [41] |
|  |  |  | 632 |  |  | Zr-O | [42] |
| 665 | 657 | 668 | 671 | 640 | 650 | Ti-O | [42] |
|  |  |  |  |  | 686 | | |
| 782 |  |  | 755 |  |  | Ti-O or Nb/Ta-O asymmetric vibration | [40,41] |
| 835 | 837 | 838 | 843 | 829 | 807 | Nb-O asymmetric stretch | [41,42] |
| 977 |  |  | 890 |  |  | Nb-O symmetric stretch | [41,42, 45,46] |
| 1004 | 1003 | 1003 |  |  |  | | |

The light response of the HEOs was investigated utilizing UV-vis spectroscopy as illustrated in Fig. 6b. This figure shows that all the samples have high light absorbance in the UV light range. However, TiZrNbTaGaO$_{10.5}$, TiZrNbTaInO$_{10.5}$ and TiZrNbTaZnO$_{10}$ also show significant absorbance in the visible light range. Herein, it is important to mention that these three samples are gray in color, as shown in the inset of Fig. 1. Gray color is usually a characteristic of oxides with oxygen vacancies, which can act as color centers and improve the light absorbance [48]. The samples containing bismuth, such as TiZrNbTaBiO$_{10.5}$ and TiNbTaGaBiO$_{10}$, do not show high absorbance in the visible region and possess a yellow appearance. The Kubelka-Munk analysis was utilized to achieve the bandgap of the materials (Fig 6c). The samples show a bandgap in a range of 2.5-2.8 eV, which is lower compared to the bandgap of TiO$_2$ anatase as a standard photocatalyst (3.2 eV) [35,49]: TiZrNbTaGaO$_{10.5}$ (2.5 eV), TiZrNbTaInO$_{10.5}$ (2.6 eV), TiZrNbTaBiO$_{10.5}$ (2.8 eV), TiZrNbTaZnO$_{10}$ (2.5 eV) and TiNbTaGaBiO$_{10}$ (2.8 eV). It should be noted that the bandgap of TiNbTaCuZnO$_9$ could not be determined due to its high light absorbance. Samples that show the second absorbance in the visible light region show another slope and energy gap in Kubelka-Munk plots. The second slope is not related to the bandgap of the second phase because it is also observed in single-phase TiZrNbTaZnO$_{10}$. The second slope should be due to the defect levels such as vacancies. The bandgap information, combined with the position for the valence band maximum obtained by XPS (Fig. 6d) allows the representation of the band strictures for all HEO as shown in Fig 6e. Considering the band positions, all samples are suitable for photocatalytic $CO_2$ photoconversion as the reaction chemical potentials lie in the center of the bandgaps.

Fig. 6f shows photoluminescence measurements, as an indicative of radiative recombinations of the photogenerated carriers. The results indicate that most of the oxides have low radiative recombination. $TiZrNbTaBiO_{10.5}$ shows a peak with a higher intensity around 610 nm which is an indication of higher recombination than the other samples. However, even the peak intensity of the photoluminescence spectrum for this material is smaller than those reported under similar measurement conditions for $TiO_2$ as a standard photocatalyst [39].

### 3.5. Photocatalytic Methanation

Fig. 7 presents the photocatalytic (a) CO, (b) $H_2$ and (c) $CH_4$ production as a function of illumination time for the six selected HEOs as well as $TiZrHfNbTaO_{11}$ and anatase $TiO_2$ for comparison. The average quantitative data for the 10 h photocatalytic test are also given in Table 4. All the synthesized HEOs produce CO, $H_2$ and $CH_4$ under the same experimental conditions, as shown in Figs. 7a-c. All the samples containing elements with $d^0 + d^{10}$ electronic configuration exhibit better methanation than the sample $TiZrHfNbTaO_{11}$ containing only $d^0$ electronic configuration. Moreover, all samples with $d^0 + d^{10}$ electronic configuration have higher methanation rate compared to anatase $TiO_2$. The sample $TiZrNbTaZnO_{10}$ performs the best among all the tested compositions for methanation followed by $TiZrNbTaBiO_{10.5}$, followed by $TiZrNbTaGaO_{10.5}$, $TiNbTaCuZnO_9$, $TiNbTaGaBiO_{10.5}$ and $TiZrNbTaInO_{10.5}$. The $CO_2$ reduction products using anatase were also included in the figures for comparison. Herein, all the HEOs show higher activity than anatase demonstrating the viability of these HEOs as photocatalysts. It should be noted that for many catalysts, both in the current study and in the literature, the CO and $CH_4$ production rates decrease over time. This behavior is commonly attributed to the availability of fresh active sites and the presence of adsorbed intermediates at the beginning of photocatalysis, leading to high activity in the first couple of hours [50–52]. However, as the reaction proceeds, surface saturation with reaction by-products, such as carbonate species or holes, can hinder the $CO_2$ conversion rate, particularly in the absence of hole scavengers [50,51].

By considering the results gathered in Fig. 7d, it is possible to examine the catalyst selectivity for $CO_2$ photoreduction and methanation. To compare the preference of each material, the selectivity for $CO_2$ conversion and methanation can be estimated by using the electron consumption rates through the following relationships.

$CO_2$ Reduction Selectivity = 100 x ($2r_{CO} + 8r_{CH4}$) / ($2r_{CO} + 8r_{CH4} + 2r_{H2}$)     (2)
Methanation Selectivity = 100 x ($8r_{CH4}$) / ($2r_{CO} + 8r_{CH4}$)     (3)

where $r(CO)$, $r(CH_4)$ and $r(H_2)$ are the production rates for CO, $CH_4$ and $H_2$ and the numbers 2 and 8 are the electrons needed for the reactions. The selectivity data are presented in Table 4 for the HEOs. All the materials have a certain preference for hydrogen production instead of $CO_2$ conversion. Since the $CO_2$ conversion experiments were performed in wet conditions, there is a constant competition between the $CO_2$ reduction reactions and the water splitting reactions [17]. Since water is everywhere in the reactor, but $CO_2$ is bubbled, the high amount of produced hydrogen is reasonable. Table 4 indicates that the HEOs with the $d^0 + d^{10}$ electronic configuration have a higher selectivity for methanation than $TiZrHfNbTaO_{11}$ with only the $d^0$ electronic configuration. Among all synthesized HEOs, TiZrNbTaZnO$_{10}$ has the highest selectivity for CO$_2$ reduction, while TiZrNbTaZnO$_{10}$ and TiZrNbTaBiO$_{10}$ exhibit the highest selectivity for methanation. The changes in selectivity are of importance because this issue has been a matter of research in many studies [2,8,16,18,19].

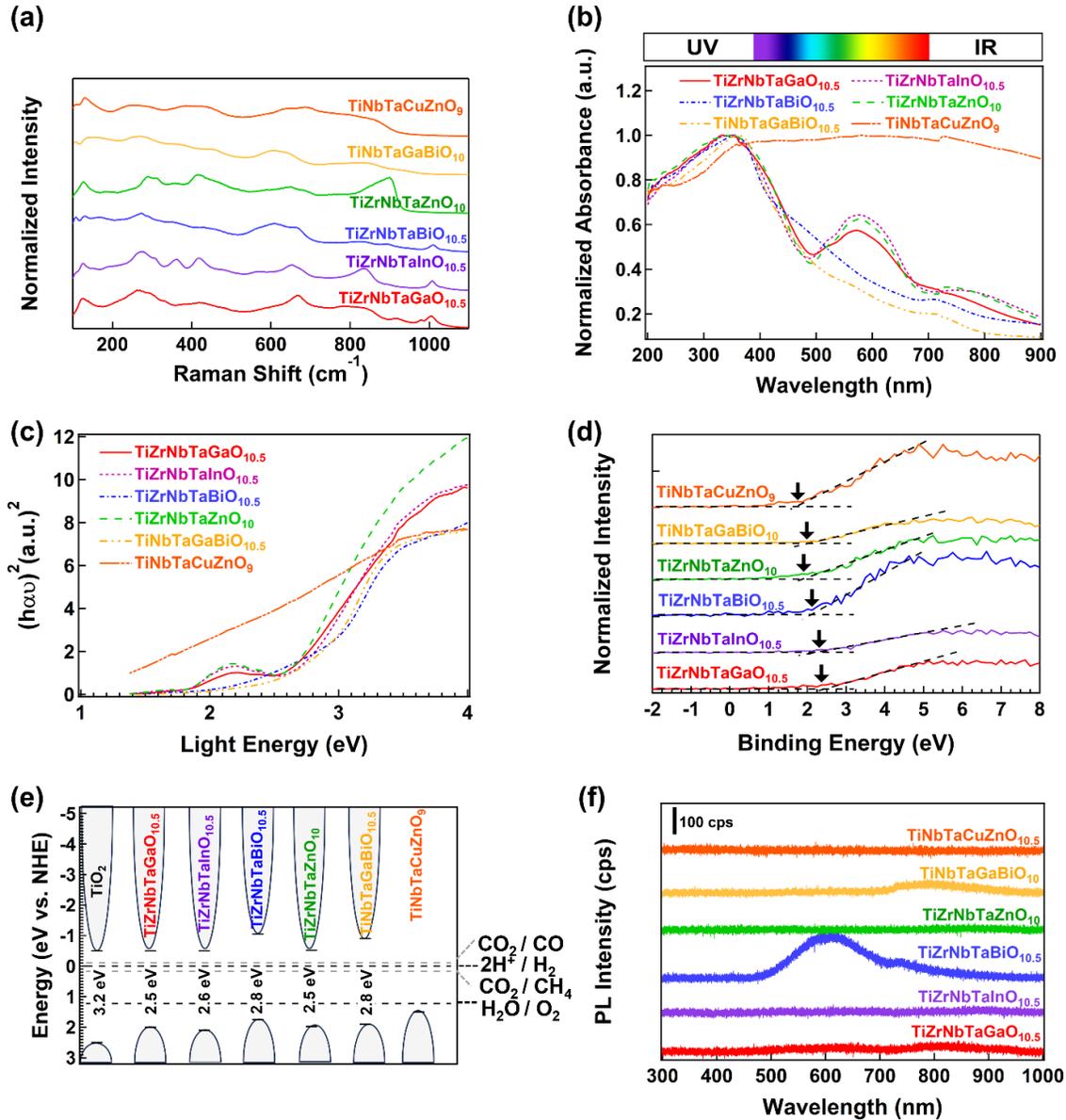

Fig. 6. Optical properties of high-entropy oxides. (a) Raman spectra, (b) light absorbance obtained utilizing UV-vis spectra, (c) Kubelka-Munk plots for calculating bandgap, (d) XPS analysis for valence band maximum calculation, (e) structure of electronic band and standard chemical potential positions for splitting water and converting CO$_2$, and (f) photoluminescence spectroscopy for radiative recombination evaluation for TiZrNbTaGaO$_{10.5}$, TiZrNbTaInO$_{10.5}$, TiZrNbTaBiO$_{10.5}$, TiZrNbTaZnO$_{10}$, TiNbTaGaBiO$_{10}$ and TiNbTaCuZnO$_9$. In Kubelka-Munk plots, $\alpha$ denotes light adsorption, h denotes Plank constant and $\upsilon$ denotes frequency of light. Data for TiO$_2$ anatase and TiZrHfNbTaO$_{10}$ were included in (e) for comparison.

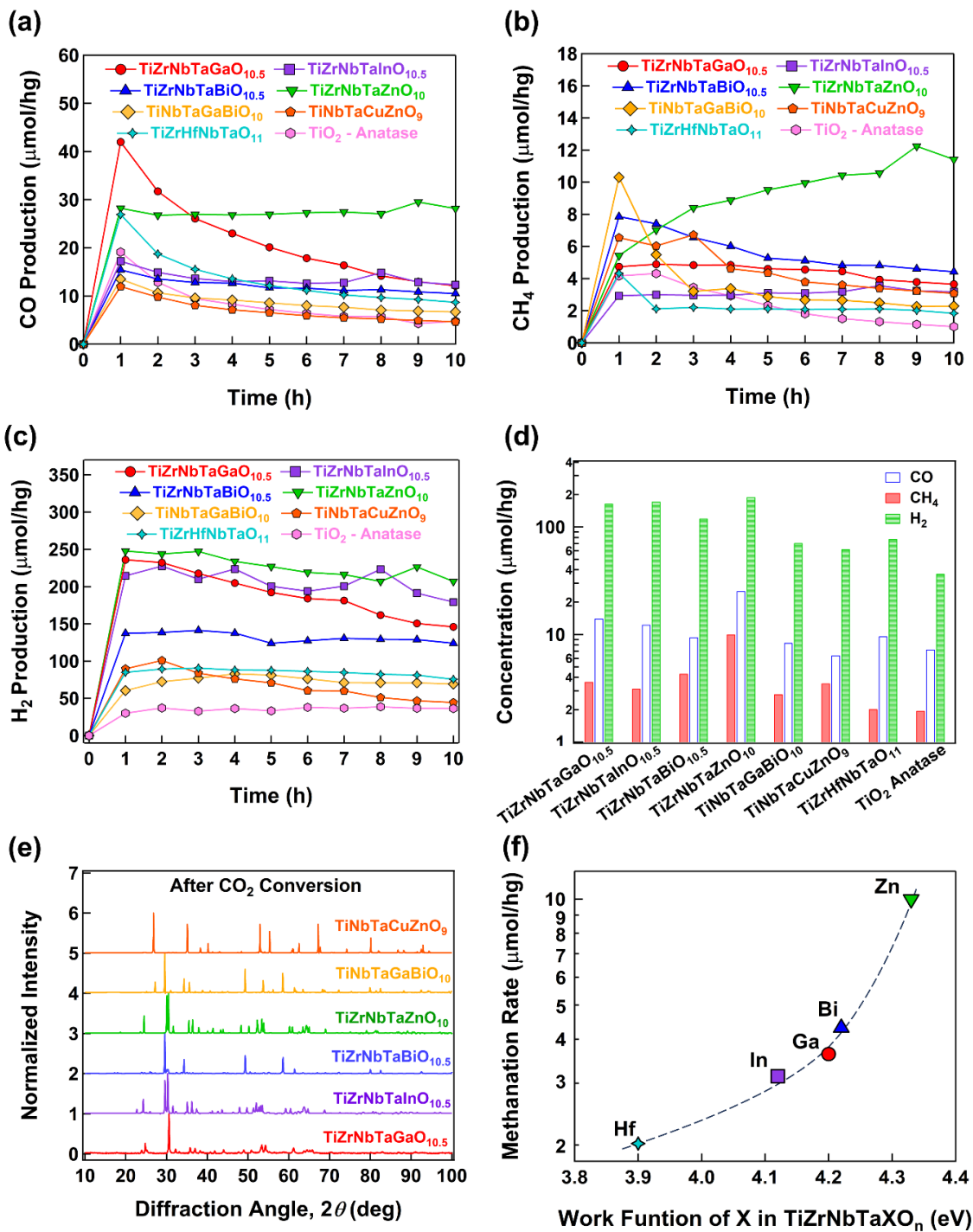

Fig. 7. Increasing the tendency for methanation in high-entropy oxides with hybrid $d^0 + d^{10}$ orbital configuration. Photocatalytic (c) $H_2$, (b) CO and (c) $CH_4$ production rates versus irradiation time, (d) average production rates within 10 h of irradiation, (e) XRD profiles of catalysts after 24 h photocatalysis for TiZrNbTaGaO$_{10.5}$, TiZrNbTaInO$_{10.5}$, TiZrNbTaBiO$_{10.5}$, TiZrNbTaZnO$_{10}$, TiNbTaGaBiO$_{10}$ and TiNbTaCuZnO$_9$, and (f) variations of methanation rate versus work function of X in TiZrNbTaXO$_n$ (X: Zn, Bi, Ga, In, Hf) oxides. Data for TiZrHfNbTaO$_{10}$ and TiO$_2$ anatase were included for comparison.

To examine the cycling stability, an extra $CO_2$ conversion test was performed using $TiZrNbTaZnO_{10}$ for 3 cycles of 24 h for each cycle. Although the measurements were conducted using a different measurement system and cannot be compared with the one given in Fig. 7, they confirmed that the material keeps 40% of its activity after 72 h. Here, it should be noted that no molecular oxygen was detected in this study for any of the catalysts (even for conventional water splitting experiments). Although $O_2$ should theoretically be produced as a by-product of water oxidation, its formation is often challenging due to the sluggish kinetics of the oxidation half-reaction and competing side reactions. Prior studies have reported that oxygen-containing intermediates, such as peroxides, carbonates or surface-bound oxygen species, can accumulate on the photocatalyst surface, preventing the release of molecular $O_2$ [51]. Additionally, the presence of oxygen vacancies and surface defects may facilitate alternative oxidation pathways, leading to the formation of bound oxygen species rather than gaseous $O_2$, as observed in similar photocatalytic systems [51,52]. In this study, without the use of a hole scavenger, photogenerated holes likely contributed to the formation of these intermediates, further complicating $O_2$ production, leading to a decrease in catalytic activity over time. To confirm that the reduced activity is not due to the instability of catalysts, the structure of the HEOs was analyzed using XRD after 24 h of photocatalytic test as shown in Fig. 7(e). The comparison of these results with the initial sample suggests that there is no phase transformation or decompositions, indicating the high stability of these HEOs.

Table 4. Electronic configuration, work function of elements added to Ti-Zr-Nb-Ta-based oxides, methanation rate, selectivity for $CO_2$ reduction and selectivity for methanation for high-entropy oxides $TiZrNbTaGaO_{10.5}$, $TiZrNbTaInO_{10.5}$, $TiZrNbTaBiO_{10.5}$, $TiZrNbTaZnO_{10}$, $TiNbTaGaBiO_{10}$ and $TiNbTaCuZnO_9$.

| | Electronic Configuration | Work function of X in $TiZrNbTaXO_{10-11}$ (eV) | Methanation Rate (µmol/h.g) | Selectivity for $CO_2$ Reduction (%) | Selectivity for Methanation (%) |
|---|---|---|---|---|---|
| **$TiZrNbTaGaO_{10.5}$** | $d^0 + d^{10}$ | Ga: 4.20 | 3.6 | 14.8 | 50.9 |
| **$TiZrNbTaInO_{10.5}$** | $d^0 + d^{10}$ | In: 4.12 | 3.1 | 12.6 | 50.5 |
| **$TiZrNbTaBiO_{10.5}$** | $d^0 + d^{10}$ | Bi: 4.32 | 4.3 | 19.7 | 64.7 |
| **$TiZrNbTaZnO_{10}$** | $d^0 + d^{10}$ | Zn: 4.33 | 9.9 | 25.6 | 61.2 |
| **$TiNbTaGaBiO_{10}$** | $d^0 + d^{10}$ | | 3.4 | 21.5 | 57.1 |
| **$TiNbTaCuZnO_9$** | $d^0 + d^{10}$ | | 3.5 | 29.5 | 68.8 |
| **$TiZrHfNbTaO_{11}$** | $d^0$ | Hf: 3.9 | 2.0 | 18.6 | 45.6 |
| **$TiO_2$ - Anatase** | $d^0$ | | 1.9 | 29.14 | 51.9 |

## 4. Discussion

HEOs have great potential for catalytic applications including photocatalysis. Lattice strain, abundant defects, and hybridized orbitals are only a few features that make them attractive for photocatalysis. However, understanding their design for different reactions still presents a challenge. Each element in high-entropy materials has a function and can drastically change the properties. In this study, several new HEOs were designed utilizing the concept of $d^0 + d^{10}$ mixed electronic configuration. Three elements (titanium, niobium and tantalum) were systematically used in each composition to understand the effect of different $d^{10}$ electronic configuration elements. Results showed higher methane formation for the mixed electronic configuration HEOs compared with the compositions only using $d^0$ electronic configuration elements ($TiZrHfNbTaO_{11}$). Table 5 compares the best high-entropy photocatalysts developed in this study with some catalysts utilized for $CO_2$ conversion in the literature [17,20-22,24,32,36,53–57]. Although the data from various articles need to be compared with care because of the differences in experimental setup, catalyst concentration, light source and so on, Table 5 demonstrates a superior methane formation (particularly per surface area of catalyst) using the high-entropy photocatalysts developed in this study. Here, the effect of the $d^{10}$ electronic configuration elements on methane formation should be discussed further.

Table 5. Comparison of $CO_2$ methanation activity of $TiZrNbTaZnO_{10}$ and $TiNbTaCuZnO_9$ as high-entropy oxides with highest methanation rate and selectivity compared with reported catalysts in literature. Photocatalyst type, photocatalyst quantity, specific surface area of photocatalysts, features of irradiation light, rate of produced $CH_4$ per photocatalyst mass and surface and selectivity towards $CH_4$ formation are given for different catalysts.

| Photocatalyst | Mass (mg) | Surface Area (m²/g) | Light Source | $CO_2$ Conversion (μmol/hg) | | Methanation (μmol/hm²) | Methanation Selectivity (%) | Reference |
|---|---|---|---|---|---|---|---|---|
| | | | | CO | $CH_4$ | $CH_4$ | | |
| $TiO_2$ | 20 | | | 1.20 | 0.38 | | 55.9 | [17] |
| $TiO_2$ | 100 | 0.18 | 300 W Hg | 2.34 | - | - | - | [36] |
| $Nb_2O_5/g - C_3N_4$ | 12 | 45.08 | 300 W Xe with 420 nm cutoff filter | 0.11 | 5.65 | 0.125 | 99.5 | [55] |
| $Cd_{1-x}Zn_xS$ ($x = 0.8$) | 45 | 119.00 | 100 W LED | 2.90 | 0.22 | 0.002 | 23.3 | [56] |
| $CuInSnS_4$ | 50 | 24.10 | 300 W Xe with 420 nm cutoff filter | | 5.83 | 0.242 | | [32] |
| $OV-α-Ga_2O_3$ | 10 | | 300W Xe | 6.3 | 3.10 | | 66.3 | [57] |
| 2% CuO - 19% $ZnO/TiO_2$ | | 43.00 | 18 W Hg vapor UVC | ~72.92 | 7.67 | 0.178 | 29.6 | [53] |
| $TiZrHfNbTaO_6N_3$ | 100 | 2.30 | 400 W Hg | 11.60 | - | | | [20] |
| $(Ga_{0.2}Cr_{0.2}Mn_{0.2}Ni_{0.2}Zn_{0.2})_3O_4$ | 20 | 16.71 | 300 W Xe | 23.01 | 2.89 | 0.173 | 33.4 | [22] |
| $Cu-(Ga_{0.2}Cr_{0.2}Mn_{0.2}Ni_{0.2}Zn_{0.2})_3O_4$ | 20 | 42.08 | 300 W Xe | 5.66 | 33.84 | 0.804 | 96.0 | [24] |
| $(NiCuMnCoZnFe)_3O_4$ | 30 | 66.48 | - | 15.89 | 8.03 | 0.121 | 66.9 | [21] |
| $(CdZnCuCoFe)S_{1.25}/ZnIn_2S_4$ | 10 | 62.20 | 300 W Xe with 420 nm cutoff | 2.43 | | | | [54] |
| $TiZrNbTaZnO_{10}$ | 100 | 0.03 | 400 W Hg | 25.30 | 9.99 | 300.8 | 61.2 | This study |
| $TiNbTaCuZnO_9$ | 100 | 0.06 | 400 W Hg | 6.37 | 3.51 | 58.58 | 68.8 | This study |

The $d^0$ electronic configuration elements behave as electron donors and the $d^{10}$ electronic configuration elements function as acceptors, and thus, their co-presence improves the charge transfer inside of the material. In addition, the cations with $d^0$ configuration and small electronegativity provide stronger adsorption sites for $CO_2$ and $H_2O$ molecules [33]. To have a view of the significance of the $d^{10}$ electronic configuration, it is useful to review earlier publications about noble metal cocatalysts. Earlier studies about noble metals suggested their ability as efficient cocatalysts, which is related to their work functions and strong carrier trapping capability [26]. For $CH_4$ production, it was observed that elements like platinum, gold and silver have good selectivity toward methane formation, and the corresponding methanation rate follows the same trend as the work function of these noble metals [18]. Therefore, it can be expected that the same concept of work function can be extended to high-entropy materials, as no cocatalyst was added to the samples but $d^{10}$ elements were included in their crystal structure in this study. With this idea in mind, it can be observed in Table 4 that there is a correlation between the work function of $d^{10}$ elements in Ti-Zr-Nb-Ta-based oxides and the methanation rate. The work functions of zinc (4.33 eV), bismuth (4.22 eV), gallium (4.20 eV), indium (4.12 eV) and hafnium (3.9 eV) [58] follow the same sequence as the methane production rate in $TiZrNbTaXO_{10-11}$ (X: Zn, Bi, Ga, In, Hf) oxides. For clarification, Fig. 7(f) plots the methanation rate versus the work function of $d^{10}$ elements. The methanation rate progressively increases with increasing the work function. A higher work function results in a better performance as shallow trapping sites for electrons, contributing to the efficient utilization of these photo-induced electrons for $CH_4$ production [18]. This feature of the $d^0 + d^{10}$ combination is also expected to suppress the charge carrier recombinations, as observed in Fig. 6f using photoluminescence. Herein, it is important to notice that the highest radiative recombination observed in this study for $TiZrNbTaBiO_{10.5}$ (based on photoluminescence intensity) is still much lower than the corresponding binary oxides [49,59–61].

Although the work function can justify the effect of $d^{10}$ cations on methanation selectivity in $TiZrNbTaXO_{10-11}$ (X: Zn, Bi, Ga, In, Hf) oxides, the situation can be more complex when two $d^{10}$ cations are added to HEOs such as in $TiNbTaGaBiO_{10}$ and $TiNbTaCuZnO_9$. In these HEOs, the absence of zirconium as a good reactant adsorption site and the synergy between the two $d^{10}$ electronic configuration elements need to be considered. The effect of combining gallium and bismuth as $d^{10}$ elements leads to a higher selectivity for $CO_2$ reduction. Even though $Ga_2O_3$ has a wide bandgap [62], the addition of $Ga^{3+}$ has shown advantages for photocatalytic hydrogen generation [63]. Conversely, the electronic configuration of bismuth includes p orbitals ($6s^2\ 6p^3$), that are highly appropriate for $CO_2$ photoreduction [64]. Depending on the oxidation state of bismuth, this $p$-band can shift and facilitate the overlapping with the p orbitals of the adsorbed $CO_2$ molecules, modifying the selectivity [64]. The oxidation state of bismuth and corresponding filled 6p orbital in $TiZrNbTaBiO_{10.5}$ might facilitate converting $CO_2$ to $CH_4$ [64]. It was reported that α-$Ga_2O_3$ has a better potential for hydrogen production than $CO_2$ conversion; however, when bismuth is added to it, the hydrogen production is inhibited due to the weak intrinsic adsorption of produced hydrogen, leading to weaker methanation [57]. This is comparable with results for $TiZrNbTaGaO_{10.5}$ in which the addition of bismuth and the formation of $TiNbTaGaBiO_{10}$ leads to weaker hydrogen production and accordingly weaker methanation.

The combination of copper and zinc is also popular in catalysts for $CO_2$ reduction reactions [65,66]. For example, adding copper and zinc to $TiO_2$ was reported to enhance the $CH_4$ production [53]. In the present study, $TiNbTaCuZnO_9$ also shows strong selectivity towards methanation from $CO_2$ conversion. Cu-based materials, specifically, can facilitate the adsorption of *CO, but their behavior depends on the oxidation states of copper [65]. The presence of multiple chemical environments around copper sites can contribute to the hydrogenation of the adsorbed *CO while other sites interact with the water molecules to form hydrogen [65]. Besides, the addition of $Zn^{2+}$ has a strong redox ability which allows it to weakly trap and release the photogenerated carriers around copper sites[67]. Moreover, the interaction of the 3d orbitals of copper and zinc and the 2p orbital of $CO_2$ was reported to enhance the electron mobility from cations to $CO_2$ molecules [21]. Improving the mobility of the carriers promotes the production of $CH_4$ because methanation requires faster kinetic compared with CO formation [22]. It should be noted that the interaction of elements is not limited to $d^{10}$ cations. For example, In-based oxides show a preference for CO production [68,69]. However, indium doping in $TiO_2$ anatase improves the selectivity towards the formation of $CH_4$ by decreasing the carrier recombination [70,71]. Still, there are many uncertainties when discussing high-entropy photocatalysts due to their lattice distortion, internal dipole moments and defects, particularly when they are synthesized by severe plastic deformation via HPT [72,73]. However, this study highlights the importance of work function and $d^0 + d^{10}$ electronic configuration for better methanation selectivity.

## 5. Conclusions

This study presents a comparative analysis of methanation using several single- and dual-phase vacancy-rich high-entropy oxides containing cations with $d^0 + d^{10}$ electronic configuration elements. The presence of $d^{10}$ electronic configuration elements improves the methanation selectivity. Moreover, the selectivity towards methanation improves by increasing the work function of $d^{10}$ cations. This work introduces an approach to adjust the composition of catalysts to achieve higher methane formation.

## Acknowledgment

JHJ thanks the Q-Energy Innovator Fellowship of Kyushu University for a scholarship. This work is funded in part by Mitsui Chemicals, Inc., Japan, in part by Grants-in-Aid from the Japan Society for the Promotion of Science (No. JP22K18737), in part by the Japan Science and Technology Agency, the Establishment of University Fellowships Towards the Creation of Science Technology Innovation (No. JPMJFS2132), in part by the ASPIRE project of the Japan Science and Technology Agency (No. JPMJAP2332), in part by the University of Rouen Normandy (No. BQRI MaP-StHy202), and in part by the CNRS Federation (No. IRMA-FR 3095).